\documentclass[sigconf]{acmart}
\usepackage[utf8]{inputenc}
\usepackage{multirow}

\usepackage{listings}
\usepackage{xcolor}

\definecolor{codegreen}{rgb}{0,0.6,0}
\definecolor{codegray}{rgb}{0.5,0.5,0.5}
\definecolor{codepurple}{rgb}{0.58,0,0.82}
\definecolor{backcolour}{rgb}{0.95,0.95,0.92}

\lstdefinestyle{mystyle}{
    backgroundcolor=\color{backcolour},   
    commentstyle=\color{codegreen},
    keywordstyle=\color{magenta},
    numberstyle=\tiny\color{codegray},
    stringstyle=\color{codepurple},
    basicstyle=\ttfamily\footnotesize,
    breakatwhitespace=false,         
    breaklines=true,                 
    captionpos=b,                    
    keepspaces=true,                 
    numbers=left,                    
    numbersep=5pt,                  
    showspaces=false,                
    showstringspaces=false,
    showtabs=false,                  
    tabsize=2
}

\definecolor{delim}{RGB}{20,105,176}
\colorlet{numb}{magenta!60!black}
\colorlet{punct}{red!60!black}
\lstdefinelanguage{json}{
    basicstyle=\small\ttfamily,
    numbers=left,
    numberstyle=\scriptsize,
    stepnumber=1,
    numbersep=8pt,
    showstringspaces=false,
    breaklines=true,
    frame=lines,
    backgroundcolor=\color{backcolour},
    literate=
     *{0}{{{\color{numb}0}}}{1}
      {1}{{{\color{numb}1}}}{1}
      {2}{{{\color{numb}2}}}{1}
      {3}{{{\color{numb}3}}}{1}
      {4}{{{\color{numb}4}}}{1}
      {5}{{{\color{numb}5}}}{1}
      {6}{{{\color{numb}6}}}{1}
      {7}{{{\color{numb}7}}}{1}
      {8}{{{\color{numb}8}}}{1}
      {9}{{{\color{numb}9}}}{1}
      {:}{{{\color{punct}{:}}}}{1}
      {,}{{{\color{punct}{,}}}}{1}
      {\{}{{{\color{delim}{\{}}}}{1}
      {\}}{{{\color{delim}{\}}}}}{1}
      {[}{{{\color{delim}{[}}}}{1}
      {]}{{{\color{delim}{]}}}}{1},
}

\title{ArchaeoDAL: A Data Lake for Archaeological Data Management and Analytics}

\author{Pengfei Liu}
\email{pengfei.liu@eric.univ-lyon2.fr}
\orcid{0000-0002-2347-3834}
\author{Sabine Loudcher}
\email{sabine.loudcher@univ-lyon2.fr}
\author{Jérôme Darmont}
\email{jerome.darmont@univ-lyon2.fr}

\affiliation{%
  \institution{Université de Lyon, Lyon 2, UR ERIC}
  \streetaddress{5 avenue Pierre Mendès France}
  \city{Lyon}
  \state{}
  \country{France}
  \postcode{69676}
}
\author{Camille Noûs}
\email{camille.nous@cogitamus.fr}
\affiliation{%
  \institution{Laboratoire Cogitamus}
  \streetaddress{5 avenue Pierre Mendès}
  \city{}
  \state{}
  \country{France}
  \postcode{69676}
}

\date{June 2021}
\copyrightyear{2021}
\acmYear{2021}
\setcopyright{acmlicensed}\acmConference[IDEAS 2021]{25th International Database Engineering \& Applications Symposium}{July 14--16, 2021}{Montreal, QC, Canada}
\acmBooktitle{25th International Database Engineering \& Applications Symposium (IDEAS 2021), July 14--16, 2021, Montreal, QC, Canada}
\acmPrice{15.00}
\acmDOI{10.1145/3472163.3472266}
\acmISBN{978-1-4503-8991-4/21/07}

\begin{abstract} 
With new emerging technologies, such as satellites and drones, archaeologists collect data over large areas. However, it becomes difficult to process such data in time. Archaeological data also have many different formats (images, texts, sensor data) and can be structured, semi-structured and unstructured. Such variety makes data difficult to collect, store, manage, search and analyze effectively. A few approaches have been proposed, but none of them covers the full data lifecycle nor provides an efficient data management system. Hence, we propose the use of a data lake to provide centralized data stores to host heterogeneous data, as well as tools for data quality checking, cleaning, transformation and analysis. In this paper, we propose a generic, flexible and complete data lake architecture. Our metadata management system exploits goldMEDAL, which is the most generic metadata model currently available. Finally, we detail the concrete implementation of this architecture dedicated to an archaeological project.

\end{abstract}



\begin{document}

\maketitle

\section{Introduction}
Over the past decade, new forms of data such as geospatial data and aerial photography have been included in archaeology research \cite{McCoy2017}, leading to new challenges such as storing massive, heterogeneous data, high-performance data processing and data governance \cite{Gattiglia2015}. As a result, archaeologists need 
a platform that can host, process, analyze and share such data.

In this context, a multidisciplinary consortium of archaeologists and computer scientists proposed the HyperThesau project\footnote{\url{https://imu.universite-lyon.fr/projet/hypertheseau-hyper-thesaurus-et-lacs-de-donnees-fouiller-la-ville-et-ses-archives-archeologiques-2018/}}, which aims at designing a data management and analysis platform. HyperThesau has two main objectives: 1) the design and implementation of an integrated platform to host, search, analyze and share archaeological data; 2) the design of an archaeological thesaurus taking the whole  data lifecycle into account, from data creation to publication. 

Classical data management solutions, i.e., databases or data warehouses, only manage previously modeled structured data (schema-on-write approach). However, archaeologists need to store data of all formats and they may discover the use of data over time. Hence, we propose the use of a data lake~\cite{Dixon2010}, i.e., a scalable, fully integrated platform that can collect, store, clean, transform and analyze data of all types, while retaining their original formats, with no predefined structure (schema-on-read approach). Our data lake, named ArchaeoDAL, provides centralized storage for heterogeneous data and data quality checking, cleaning, transformation and analysis tools. Moreover, by including machine learning frameworks into ArchaeoDAL, we can achieve descriptive and predictive analyses. 

Many existing data lake solutions provide architecture and/or implementation, but few include a metadata management system, which is nevertheless essential to avoid building a so-called data swamp, i.e., an unexploitable data lake \cite{Inmon2016,Sawadogo2019}. Moreover, none of the existing metadata management systems can provide all the needed metadata features we need. For example, in archaeology, thesauri are often used for organizing and searching data. Therefore, the metadata system must allow users to define one or more thesauri, associate data with specific terms and create relations between terms, e.g., synonyms and antonyms. Thus, we conclude that existing data lake architectures, including metadata systems, are not generic, flexible and complete enough for our purpose. 

To address these problems, we propose in this paper a \textit{generic}, \textit{flexible} and \textit{complete} data lake architecture. Moreover, our metadata model exploits and enriches
goldMEDAL, which is
the most generic metadata model currently available  \cite{dolap2021}. To illustrate the flexibility and completeness of ArchaeoDAL's architecture, we provide a concrete implementation dedicated to the HyperThesau project. With a fully integrated metadata management and security system, we can not only ensure data security, but also track all data transformations.

The remainder of this paper is organized as follows. In Section~\ref{sec:Background}, we review and discuss existing data lake architectures. In Section~\ref{sec:Abstract_arch_imp_dep}, we present ArchaeoDAL's abstract architecture, implementation and deployment. In Section~\ref{sec:application_example}, we present two archaeological application examples. In Section~\ref{sec:Conclusion}, we finally conclude this paper and present future works.

\section{Data Lake Architectures}
\label{sec:Background}

The concept of data lake was first introduced by Dixon \cite{Dixon2010} in association with the Hadoop file system, which can host large heterogeneous data sets without any predefined schema. Soon after, the data lake concept was quickly adopted \cite{Fang2015,Oleary2014}. With the growing popularity of data lakes, many solutions were proposed. After studying them, we divide data lake architectures into two categories: 1)~data storage-centric architecture; 2)~data storage and processing-centric architecture. 

\subsection{Data Storage-Centric Architectures}

In the early days, a data lake was viewed as a central, physical storage repository 
for any type of raw data, 
aiming for future insight. 
In this line, Inmon proposes an architecture that organizes data by formats, in so-called data ponds~\cite{Inmon2016}. The raw data pond is the place where data first enters the lake. The analog data pond stores data generated by sensors or machines. The application data pond stores data generated by applications. Finally, the textual data pond stores unstructured, textual data.

Based on such zone solutions, Gorelik proposes that a common data lake architecture includes four zones~\cite{Gorelik2019}: a landing zone that hosts raw ingested data; a gold zone that hosts cleansed and enriched data; a work zone that hosts transformed, structured data for analysis; and a sensitive zone that hosts confidential data. Bird also proposes a similar architecture~\cite{Ian2019}. Such architectures organize data with respect to how deeply data are processed and security levels. 

The advantage of storage-centric architectures is that they provide a way to organize data inside a data lake by default. However, the predefined data organization may not satisfy the requirements of all projects. For example, HyperThesau needs to store data from different research entities. Thus, one requirement is to organize data by research entities first. Yet, the bigger problem of storage-centric architectures is that they omit important parts of a data lake, e.g., data processing, metadata management, etc.

\subsection{Data Storage and Processing-Centric Architectures}

With the evolution of data lakes, they have been viewed as platforms, resulting in more complete architectures.
Alrehamy and Walker propose a ``Personal Data Lake'' architecture that consists of five components~\cite{Alrehamy2015}, which addresses data ingestion and metadata management issues. However, it transforms data into a special JSON object that stores both data and metadata. By changing the original data format, this solution contradicts the data lake philosophy of conserving original data formats. 

\begin{figure}[hbt]
\centering
\includegraphics[width=8.5cm]{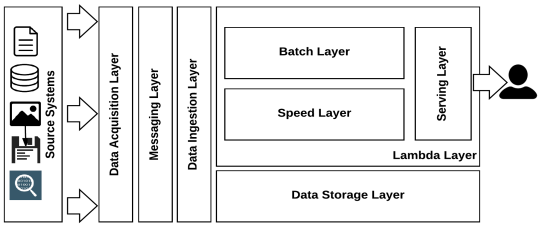}
\caption{Lambda data lake architecture \cite{John2017}}
\label{fig:datalake_john}
\end{figure}

Pankaj and Tomcy propose a data lake architecture based on the Lambda architecture (Figure~\ref{fig:datalake_john}) that covers almost all key stages of the data life cycle~\cite{John2017}. However, it omits  metadata management and security issues. Moreover, not all data lakes need near real-time data processing capacities. 

Mehmood et al. propose an interesting architecture consisting of four layers: a data ingestion layer that acquires data for storage; a data storage layer; a data exploration and analysis layer; a data visualization layer~\cite{Mehmood2019}. This architecture is close to our definition of a data lake, i.e., a fully integrated platform to collect, store, transform and analyze data for knowledge extraction. Moreover, Mehmood et al. provide an implementation of their architecture. However, although they mention the importance of metadata management, they do not include a metadata system in their architecture. Eventually, data security is not addressed and data visualization is the only proposed analysis method. Raju et al. propose an architecture that is similar to Mehmood et al.'s~\cite{Raju2018}. They essentially use a different tool-set to implement their approach and also omit to take metadata management and data security into account. 

\subsection{Discussion}
\label{sec:datalake_evaluation}
In our opinion, a data lake architecture must be \textit{generic}, \textit{flexible} and \textit{complete}. Genericity implies that the architecture must not rely on any specific tools nor frameworks. Flexibility means that users must be able to define their own ways of organizing data. Completeness means that not only functional features (e.g., ingestion, storage, analysis, etc.) must be handled, but also non-functional features (e.g., data governance and data security). Table~\ref{tab:archi_eval} provides an evaluation of seven data lake architectures with respect to these three properties.

\begin{table}[hbt]
  \caption{Comparison of data lake architectures}
  \label{tab:archi_eval}
  \begin{tabular}{lccc}
    \toprule
    \textbf{Architecture} & Generic & Flexible & Complete\\
    \midrule
    Alrehamy and Walker (2015)& \checkmark & &\checkmark \\
    Inmon (2016) & \checkmark & &\\
    Pankaj and Tomcy (2017)& \checkmark & \checkmark & \\
    Raju et al. (2018)& &\checkmark&\\
    Bird (2019)& \checkmark & & \\
    Gorelik (2019) & \checkmark && \\
    Mehmood et al. (2019)&&\checkmark& \\
  \bottomrule
\end{tabular}
\end{table}

The solutions by Mehmood et al. and Raju et al. are not generic because their architecture heavily relies on certain tools. The zone architectures by Inmon, Gorelik and Bird are not flexible, because they force a specific data organization. Finally, Alrehamy and Walker's platform is the only complete architecture that addresses data governance and security, but is not a canonical data lake.

\section{ArchaeoDAL's Architecture, Implementation and Deployment}
\label{sec:Abstract_arch_imp_dep}
\begin{figure}[hbt]
\centering
\includegraphics[width=8.5cm]{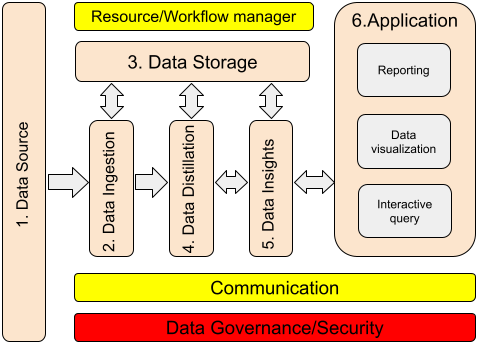}
\caption{ArchaeoDAL's architecture}
\label{fig:abs_archi}
\end{figure}

In this section, we propose a generic, flexible abstract data lake architecture that covers the full data lifecycle (Figure~\ref{fig:abs_archi}) and contains eleven layers. The orange layers (from layer 1 to layer 6) cover the full data lifecycle. After data processing in these layers, data become clearer and easier to use for end-users. The yellow layers cover non-functional requirements. After the definition of each layer, we present how each layer is implemented in our current ArchaeoDAL instance. As this instance is dedicated to the project HyperThesau, it does not cover all the features of the abstract architecture. For example, real-time data ingestion and processing are not implemented. However,real-time or near real-time data ingestion and processing feature can be achieved by adding tools such as Apache Storm\footnote{\url{https://storm.apache.org/}} or Apache Kafka\footnote{\url{https://kafka.apache.org/}} in the data ingestion and data insights layers.

\label{sec:archi_description}

\subsection{Data Source Layer} 
In the data source layer, we gather the basic properties of data sources, e.g., volume, format, velocity, connectivity, etc. Based on these properties, data engineers can determine how to import data into the lake. If metadata are required, data engineers must also  find the best fitting metadata model to govern input data. 

In our instance of ArchaeoDAL, we have data sources such as relational databases and various files stored in the archaeologists’ personal computers.

\subsection{Data Ingestion Layer}
The data ingestion layer provides a set of tools that allow users to perform batch or real-time data ingestion. Based on the data source properties that are gathered in the data source layer, data engineers can choose the right tools and plans to ingest data into the  lake. They must also consider the capacity of the data lake to avoid data loss, especially for real-time data ingestion. During ingestion, metadata  provided by the data sources, e.g., the name of excavation sites or instruments, must be gathered as much as possible. After data are loaded into the lake, we may lose track of  data sources. It is indeed more difficult to gather this kind of metadata without knowledge about data sources.

ArchaeoDAL's implementation exploits Apache Sqoop\footnote{\url{https://sqoop.apache.org/}} to ingest structured data and Apache Flume\footnote{\url{https://flume.apache.org/}} to ingest semi-structured and unstructured data. Apache Sqoop can efficiently transfer bulk data from structured data stores such as relational databases. Apache Flume is a distributed service for efficiently collecting, aggregating and moving large amounts of data. For one-time data ingestion, our instance provides Sqoop scripts and a web interface to ingest bulk data. For repeated data loading, we developed Flume agents to achieve automated data ingestion.  

\subsection{Data Storage Layer}
The data storage layer is the core layer of a data lake. It must have the capacity to store all data, e.g., structured, semi-structured and unstructured data, in any format. 

ArchaeoDAL's implementation uses the Hadoop Distributed File System\footnote{\url{https://hadoop.apache.org/}} (HDFS) to store ingested data, because HDFS stores data on commodity machines and provides horizontal scalability and fault tolerance. As a result, we do not need to build large clusters. We just add nodes when data volume grows. 
To better support the storage of structured and semi-structured data, we add two tools: Apache Hive\footnote{\url{https://hive.apache.org/}} to store data with explicit data structures and Apache HBase\footnote{\url{https://hbase.apache.org/}} that is a distributed, versioned, column-oriented database that provides better semi-structured data retrieval speed.

\subsection{Data Distillation Layer}
The data distillation layer provides a set of tools for data cleaning and encoding formalization. Data cleaning refers to eliminating errors such as duplicates and type violations, e.g., a numeric column contains non-numeric values. Data encoding formalization refers to converting various data and character encoding, e.g., ASCII, ISO-8859-15, or Latin-1, into a unified encoding, e.g., UTF-8, which covers all language symbols and graphic characters. 

ArchaeoDAL's implementation uses Apache Spark\footnote{\url{https://spark.apache.org/}} to clean and transform data.
We developed a set of Spark programs that can detect duplicates, NULL values and type violations. Based on the percentage of detected errors, we can hint at data quality.

\subsection{Data Insights Layer} 
The data insights layer provides a set of tools for data transformation and exploratory analysis. Data transformation refers to transforming data from one or diverse sources into specific models that are ready to use for data application, e.g., reporting and visualization. Exploratory data analysis refers to the process of performing initial investigations on data to discover patterns, test hypotheses, eliminate meaningless columns, etc. Transformed data may also be persisted in the data storage layer for later reuse.

ArchaeoDAL's implementation resorts to Apache Spark to perform data transformation and exploratory data analysis. Spark also provides machine learning libraries that allow developers to perform more sophisticated exploratory data analyses.

\subsection{Data Application Layer}
The data application layer provides applications that allow users to extract value from data. For example, a data lake may provide an interactive query system to do descriptive and predictive analytics. It may also provide tools to produce reports and visualize data.

In ArchaeoDAL's implementation, we use a Web-based notebook, Apache Zeppelin\footnote{\url{https://zeppelin.apache.org/}}, as the front end. Zeppelin connects to the data analytics engine Apache Spark that can run a complex directed acyclic graph of tasks for processing data. Our notebook interface supports various languages and their associated Application Programming Interfaces (APIs), e.g., R, Python, Java, Scala and SQL. It provides a default data visualization system that can be enriched by Python or R libraries.

ArchaeoDAL also provides a web interface that helps users download, upload or delete data.

\subsection{Data Governance Layer}
The data governance layer provides a set of tools to establish and execute plans and programs for data quality control \cite{Vijay2010}. This layer is closely linked to the data storage, ingestion, distillation, insights and application layers to capture all relevant metadata. A key component of the data governance layer is a metadata model~\cite{Sawadogo2019}. 

\subsubsection{Metadata Model}
\label{Metadata_model}
\begin{figure}[hbt]
            \centering
            \includegraphics[width=8.5cm]{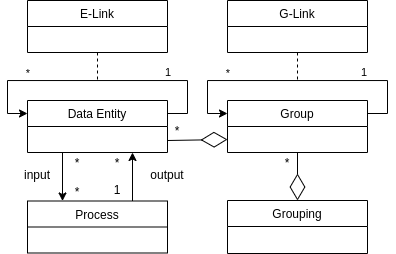}
            \caption{goldMEDAL's concepts \cite{dolap2021}}
            \label{fig:goldmedal}
        \end{figure}

In ArchaeoDAL, we adopt goldMEDAL~\cite{dolap2021}, which is modeled at the conceptual (formal description), logical (graph model) and physical (various implementations) levels. goldMEDAL features four main metadata concepts (Figure~\ref{fig:goldmedal}): 1)~\textit{data entities}, i.e., basic data units such as spreadsheet tables or textual documents; 2)~\textit{groupings} that bring together data entities w.r.t. common properties in \textit{groups}; 3)~\textit{links} that associate either data entities or groups with each other; and 4)~\textit{processes}, i.e., transformations applied to data entities that produce new data entities. All concepts bear metadata, which make goldMEDAL the most generic metadata model in the literature, to the best of our knowledge.

However, we encountered a problem of terminology variation when creating metadata. goldMEDAL does indeed not provide explicit guidance for metadata creation. This can lead to consistency and efficiency problems. For example, when users create data entities, they have their own way of defining attributes, i.e., key/value pairs that describe the basic properties of data. Without a universal guideline or template, every user can invent his own way. The number, name and type of attributes can be different. As a result, without explicit guidelines or templates, it may become quite difficult to retrieve or search metadata.

Thus, we enrich goldMEDAL with a new concept, \textit{data entity type}, which explicitly defines the number, names and types of the attributes in a data entity. 

All data entity types form a type system that specifies how metadata describe data inside the lake. Since each and every data lake has specific requirements on how to represent data to fulfill domain-specific requirements, metadata must contain adequate attributes. Thus, we need to design a domain-specific type system for each domain. For example, in the HyperThesau project, users need not only semantic metadata to understand the content of data, but also geographical metadata to know where archaeological objects are discovered. As a result, the type system is quite different from other domains. 

In sum, the benefits of having a data entity type system include: 1)~\textit{consistency}, a universal definition of metadata can avoid terminology variations that may cause data retrieval problems; 2)~\textit{flexibility}, a domain-specific type system helps define specific metadata for requirements in each use case; 3)~\textit{efficiency}, with a given metadata type system, it is easy to write and implement search queries. Because we know in advance the names and types of all metadata attributes, we can filter data with metadata predicates such as $upload\_date > 10/02/2016$.

\subsubsection{Thesaurus Modeling}
\label{sec:thesaurus}
Although thesauri, ontologies and taxonomies can definitely be modeled with goldMEDAL,  its specifications do not provide details on how to conceptually and logically model such semantic resources, while we especially need to manage thesauri in the HyperThesau project.

A thesaurus consists of a set of \textit{categories} and \textit{terms} that help regroup data. A category may have one and only one parent. A category without a parent is called the root category. A category may have one or more children. The child of a category is called \textit{subcategory} or \textit{term}. A term is a special type of category that has no child but must have a parent category. A term may  have relationships with other terms (related words, synonyms, antonyms, etc.). 

Fortunately, categories and terms can easily be modeled as data entities and structured with labeled links, with labels defined as link metadata.
It is also easy to extend this thesaurus model to represent ontologies or taxonomies.

\subsubsection{Data Governance Implementation}
\label{sec:data_govern_implementation}
The HyperThesau project does not require sophisticated data governance tools to fix decision domains and decide who takes decisions to ensure effective data management~\cite{Vijay2010}. Thus, ArchaeoDAL's data governance layer implementation only focuses on how to use metadata to govern data inside the lake. We use Apache Atlas\footnote{\url{https://atlas.apache.org/}}, which is a data governance and metadata management framework, to implement our extended version of goldMEDAL (Section~\ref{Metadata_model}). With these metadata, we build a data catalog that allows searching and filtering data through different metadata attributes, organize data with user-defined classifications and the thesaurus and trace data lineage.

\subsection{Data Security Layer}
\label{sec:security}
The data security layer provides tools to ensure data security. It should ensure the user's authenticity, data confidentiality and integrity. 

ArchaeoDAL's implementation orchestrates more than twenty tools and frameworks, most of which have their own authentication system. If we used their default authentication systems, a given user could have twenty login and password pairs. But even if a user decided to use the same login and password for all services, s/he would have to change it twenty times in case of need. To avoid this, we have deployed an OpenLDAP server as a centralized authentication server. All services connect to the centralized authentication server to check user login and password. The access control system consists of two parts. First, we need to control the access to each service. For example, when a user wants to create a new thesaurus, Atlas needs to check whether this user has the right to. Second, we need to control the access to data in the lake. A user may access data via different tools and the authorization answer of these tools must be uniform. We use Apache Ranger\footnote{\url{https://ranger.apache.org/}} to set security policy rules for each service. For data access control, we implement a role-based access control (RBAC) system by using the Hadoop group mapping service.

Data security is enforced by combining the three systems. For example, a user wants to view data from a table stored in Hive. S/he uses his/her login and password to connect to the Zeppelin notebook. This login and password are checked by the OpenLDAP server. After login, s/he runs a SQL query that is then submitted to Hive. Before query execution, Hive  sends an authorization request to Ranger. If Ranger allows the user to run the query, it starts and retrieves data with the associated user credentials. Then, HDFS checks whether the associated user credentials have the right to access data. If a user does not have the right to read data, an access denied exception is produced.

\subsection{Workflow Manager Layer}
The workflow manager layer provides tools to automate the flow of data processes. This layer is optional. 


For now, we have not identified any repeated data processing tasks in the HyperThesau project. As a result, we have not implemented this layer. However, it can easily be implemented by integrating tools such as Apache Airflow\footnote{\url{https://airflow.apache.org/}} or Apache NiFi\footnote{\url{https://nifi.apache.org/}}. 

\subsection{Resource Manager Layer}
As data lakes may involve many servers working together, the resource manager layer is responsible for negotiating resources and scheduling tasks on each server. This layer is optional, because a reasonably small data lake may be implemented on one server only.


As ArchaeoDAL rests on a distributed storage and computation framework, a resource manager layer is mandatory. We use YARN\footnote{\url{https://yarnpkg.com/}}, since this resource manager can not only manage the resources in a cluster, but also schedule jobs. 

\subsection{Communication Layer}
The communication layer provides tools that allow other layers, e.g., data application, data security and data governance, to communicate with each other. It must provide both synchronous and asynchronous communication capability. For example, a data transformation generates new metadata that are registered by the data governance layer. Metadata registration should not block the data transformation process. Therefore, the data insights and governance layers require asynchronous communication. However, when a user wants to visualize data, the data application and security layers require synchronous communication, since it is not desirable to have a user read data before the security layer authorizes the access. 

ArchaeoDAL's implementation uses Apache Kafka, which provides a unified, high-throughput, low-latency platform for handling real-time data feeds. Kafka can connect by default many frameworks and tools, e.g., Sqoop, Flume, Hive and Spark. It also provides both synchronous and asynchronous communication.

\subsection{ArchaeoDAL's Deployment}
We have deployed ArchaeoDAL's implementation in a self-hosted cloud. The current platform is a cluster containing 6 virtual machines, each having 4 virtual cores, 16 GB of RAM and 1 TB of disk space. We use Ambari\footnote{\url{https://ambari.apache.org/}} to manage and monitor the virtual machines and installed tools. The current platform allows users to ingest, store, clean, transform, analyze, visualize and share data by using different tools and frameworks.  

ArchaeodAL already hosts the data of two archaeological research facilities, i.e.,  Bibracte\footnote{\url{http://www.bibracte.fr/}} and Artefacts\footnote{\url{https://artefacts.mom.fr/}}. Artefacts currently amounts to 20,475 records and 180,478 inventoried archaeological objects. Bibracte currently contains 114 excavation site reports that contain 30,106 excavation units and 83,328 inventoried objects. We have imported a thesaurus developed by our partner researchers from the \textit{Maison de l'Orient et de la Méditerranée}\footnote{\url{https://www.mom.fr/}}, who study ancient societies in all their aspects, from prehistory to the medieval world, in the Mediterranean countries, the Near and Middle-East territories. This thesaurus implements the ISO-25964 norm. A dedicated thesaurus by the linguistic expert of the HyperThesau project is also in the pipe. With the help of our metadata management system, users can associate data with the imported thesaurus, and our search engine allows users to search and filter data based on the thesaurus.

\section{Application Examples}
\label{sec:application_example}

In this section, we illustrate how ArchaeoDAL supports users throughout the archaeological data lifecycle, via metadata. We also demonstrate the flexibility and completeness of ArchaeoDAL (Section~\ref{sec:datalake_evaluation}). 

\subsection{Heterogeneous Archaeological Data Analysis}

In this first example, we show how to analyze heterogeneous archaeological data, how to generate relevant metadata during data ingestion and transformation, and how to organize data flexibly via the metadata management system.

The Artefacts dataset consists of a SQL database of 32 tables and a set of files that stores detailed object descriptions as semi-structured data. This dataset inventories 180,478 objects.

The data management system is implemented with Apache Atlas (Section~\ref{sec:data_govern_implementation}) and provides three ways to ingest metadata: 1) pre-coded atlas hook (script), 2) self-developed atlas hook and 3) REpresentational State Transfer (REST) API.

\subsubsection{Structured Data Ingestion and Metadata Generation}

\begin{figure*}[hbt]
\centering
\includegraphics[width=14.5cm]{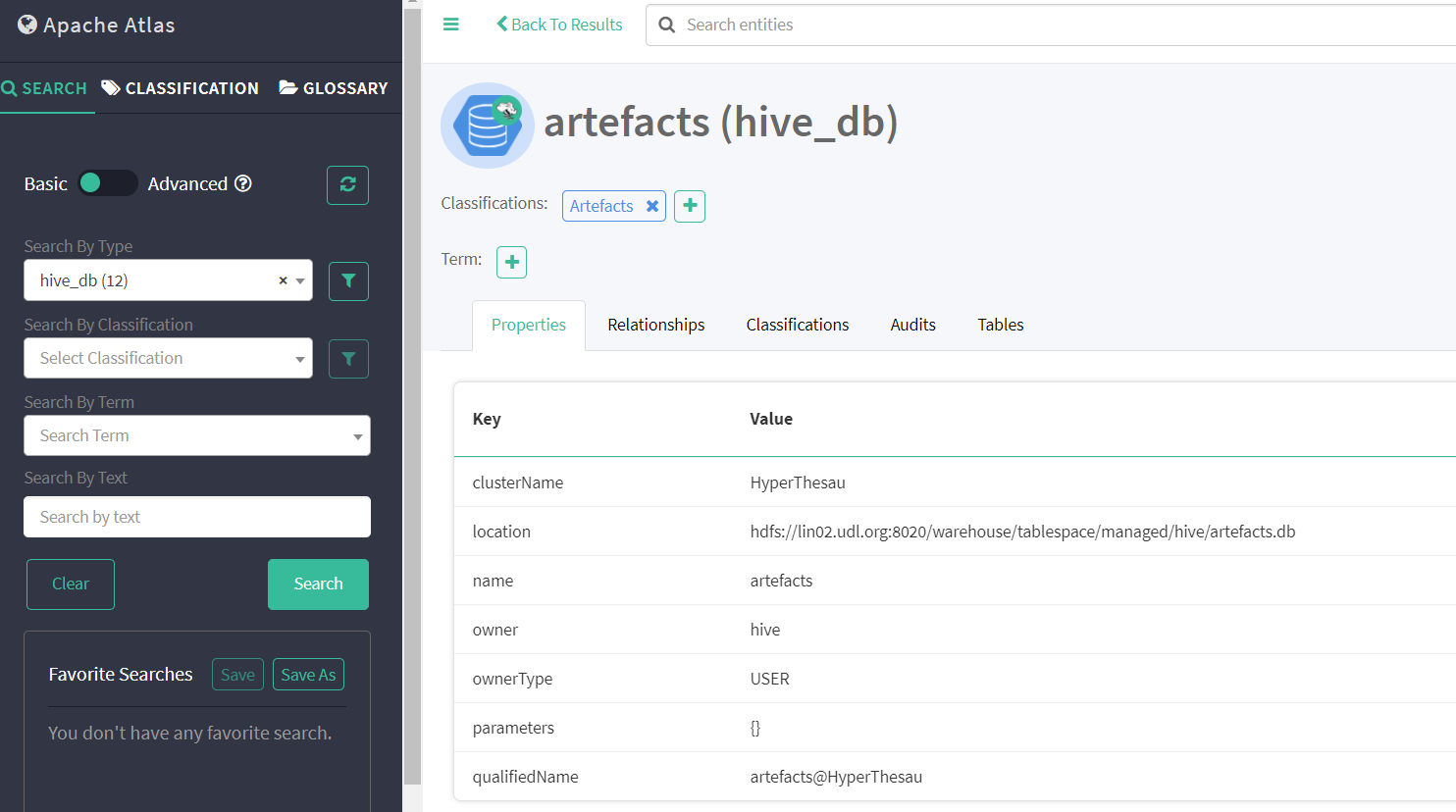}
\caption{Database metadata}
\label{fig:artefacts_db}
\end{figure*}

\begin{figure*}[hbt]
\centering
\includegraphics[width=14.5cm]{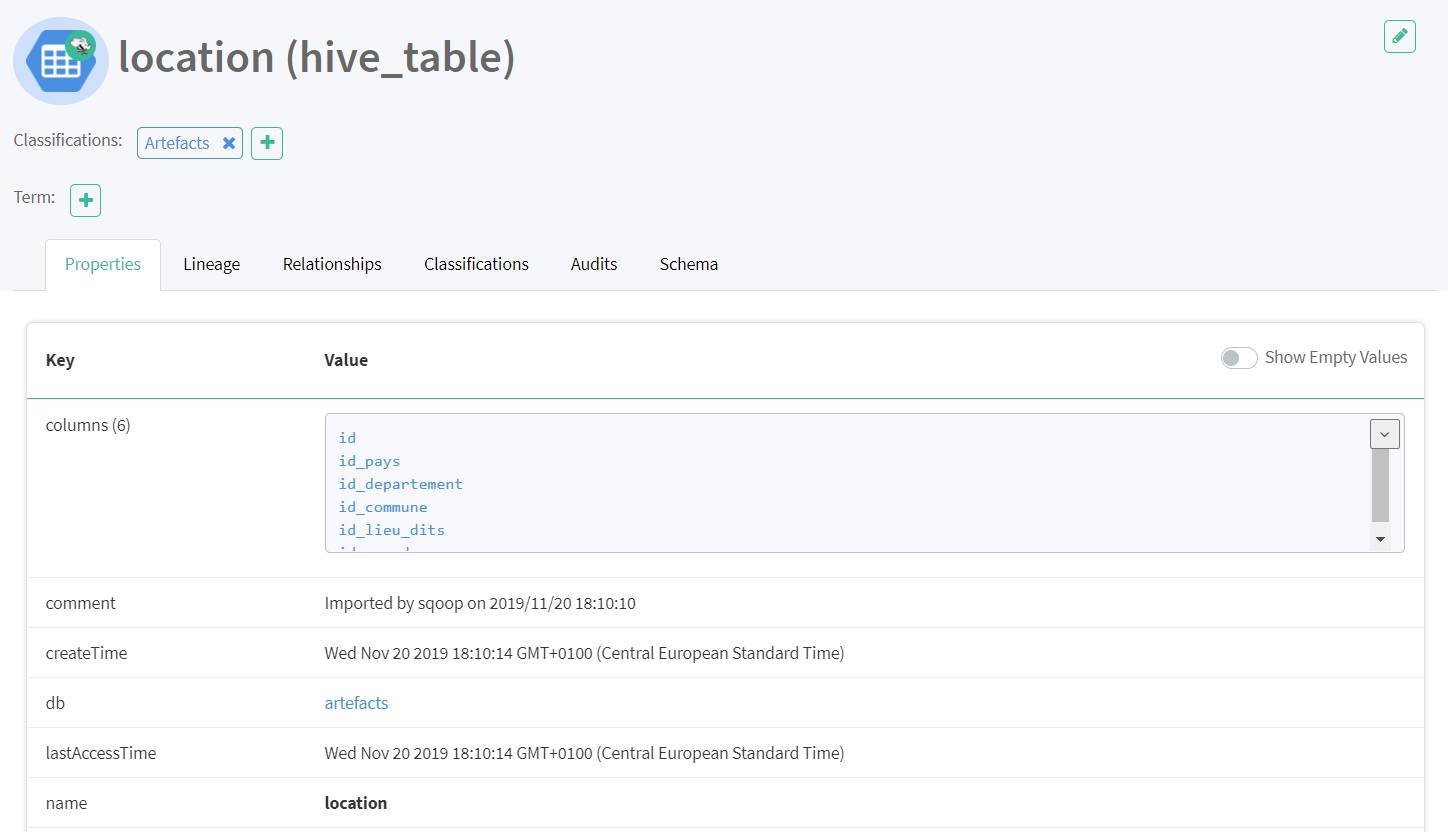}
\caption{Table metadata}
\label{fig:artefacts_table}
\end{figure*}

To import data from a SQL database, we use a hook dedicated to Sqoop that can generate the metadata of the imported data and ingest them automatically. A new database is created in ArchaeoDAL and its metadata is generated and inserted into the Atlas instance (Figure~\ref{fig:artefacts_db}). Figure~\ref{fig:artefacts_table} shows an example of table metadata inside Artefacts.

\subsubsection{Semi-structured Data Ingestion and Metadata Generation}
\begin{figure*}[hbt]
\centering
\includegraphics[width=14.5cm]{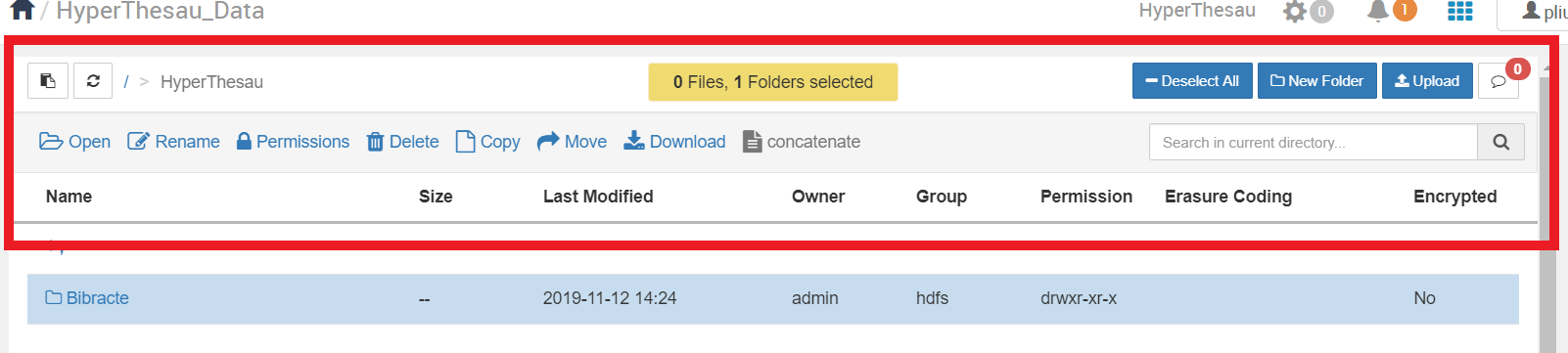}
\caption{Data upload interface}
\label{fig:file_upload}
\end{figure*}

 We provide three ways to ingest semi-structured data. The simplest way is to use Ambari's Web interface (Figure~\ref{fig:file_upload}). The second way is to use the HDFS command-line client.

The third way is to use a data ingestion tool. ArchaeoDAL provides a tool called Flume. The Flume agent can monitor any file system of any computer. When a file is created on the monitored file system, the Flume agent will upload it to ArchaeoDAL automatically. 

\begin{figure*}[hbt]
\centering
\includegraphics[width=14.5cm]{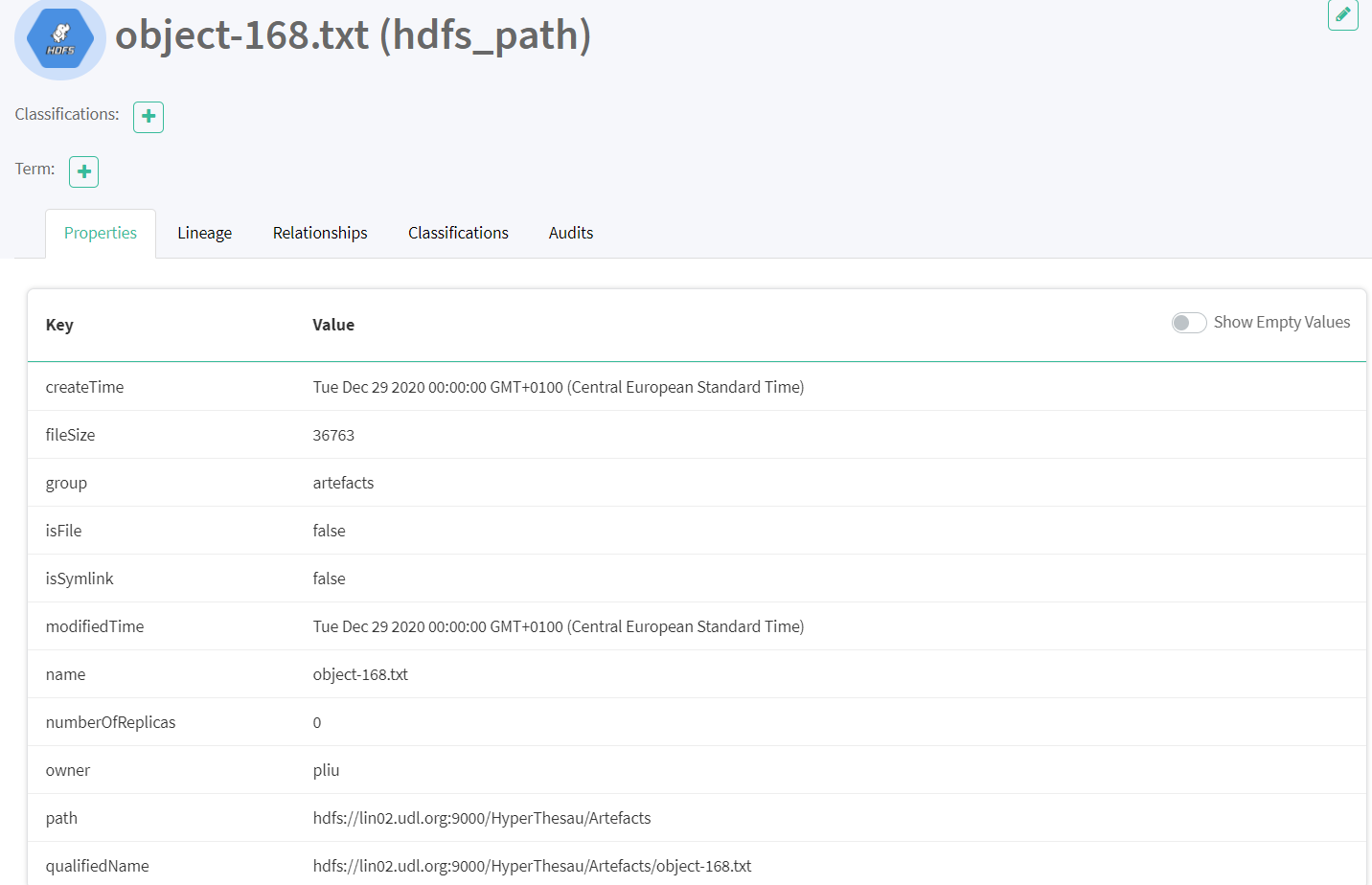}
\caption{Sample metadata visualization in Altas}
\label{fig:hdfs_hook}
\end{figure*}

Now that we have uploaded the required files into ArchaeoDAL, we need to generate  and insert the metadata into Atlas. Since Atlas does not provide a hook for HDFS, we develop our own\footnote{https://github.com/pengfei99/AtlasHDFSHook}. This hook is triggered by the HDFS create, update and delete events. For example, the upload action 
generates a file creation event in HDFS, which in turn triggers our metadata generation hook (Listing~\ref{lst:hdfs_meta}).
After the hook has uploaded metadata into Atlas, it can be visualized (Figure~\ref{fig:hdfs_hook}).

\begin{lstlisting}[language=JSON,caption={Sample generated metadata from a HDFS file},label=lst:hdfs_meta]
{ "entities": [ {  
      "typeName": "hdfs_path",
      "createdBy": "pliu",
      "attributes": {
        "qualifiedName": "hdfs://lin02.udl.org:9000/HyperThesau/Artefacts/object-168.txt",
        "name": "object-168.txt",
        "path": "hdfs://lin02.udl.org:9000/HyperThesau/Artefacts",
        "user":"pliu",
        "group":"artefacts",
        "creation_time":"2020-12-29",
        "owner":"pliu",
        "numberOfReplicas":0,
        "fileSize":36763,
        "isFile":true
      } }]}
\end{lstlisting}

We also developed an Atlas API in Python\footnote{\url{https://pypi.org/project/atlaspyapi/}} to allow data engineers to generate and insert metadata into Atlas more easily. As Amazon S3 is the most popular cloud storage, we also developed an Atlas S3 hook\footnote{\url{https://pypi.org/project/atlass3hook/}}. 

\subsubsection{Data Transformation and Metadata Generation for Data Lineage Tracing}
Once Artefacts data are ingested into ArchaeoDAL, let us imagine that an archaeologist wants to link the detailed description of objects (stored in table \textit{objects}) with their discovery and storage locations. 
The first step is to convert the semi-structured object descriptions into structured data. We developed a simple Spark Extract, Transform and Load (ETL) script for this sake. Then, we save the output structured data into Hive tables \textit{location} (discovery location) and \textit{musee} (the museum where objects are stored).

Eventually, we join the three tables into a new table called \textit{objects\_origin} that contains the objects' descriptions and their discovery and storage locations.

\begin{figure*}[hbt]
\centering
\includegraphics[width=14.5cm]{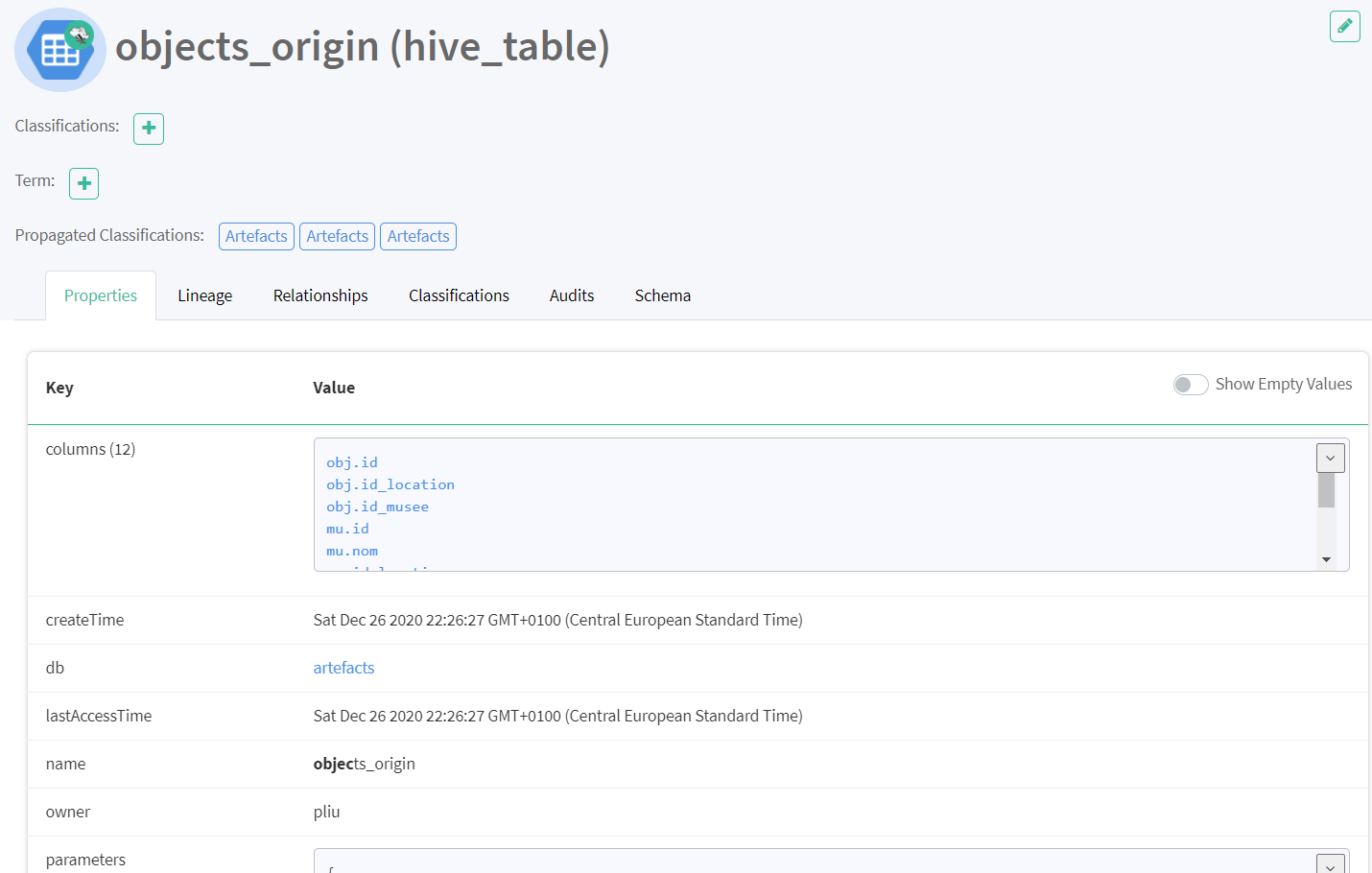}
\caption{Metadata \textit{per se}}
\label{fig:artefact_obj_origin}
\end{figure*}

\begin{figure*}[hbt]
\centering
\includegraphics[width=14.5cm]{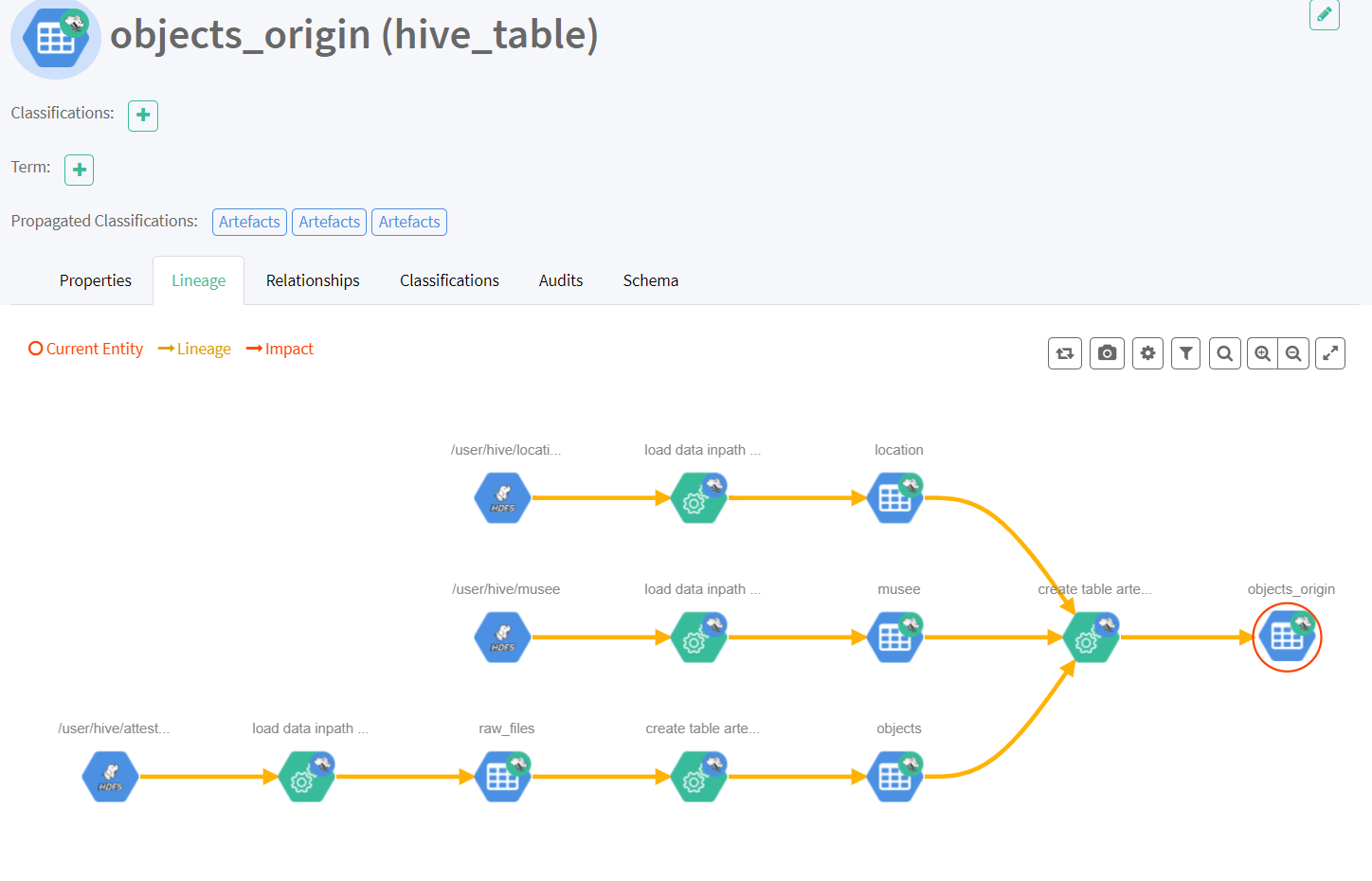}
\caption{Lineage}
\label{fig:artefact_lineage}
\end{figure*}

Thereafter, \textit{objects\_origin}'s metadata can be gathered into Atlas with the help of the default Hive hook\footnote{\url{https://atlas.apache.org/1.2.0/Hook-Hive.html}} and a Spark hook developed by Hortonworks\footnote{\url{https://github.com/hortonworks-spark/spark-atlas-connector}}. All Hive and Spark data transformations are tracked and all relevant metadata are pushed automatically into Atlas.
Figures~\ref{fig:artefact_obj_origin} and~\ref{fig:artefact_lineage} show table \textit{objects\_origin}'s metadata and lineage, respectively.

\subsubsection{Flexible Data Organization}
\label{sec:flexible-data-org}
As we mentioned in Section~\ref{sec:datalake_evaluation}, existing data lake solutions do not allow users to define their own ways of organizing data, while ArchaeoDAL users should. Moreover, ArchaeoDAL must allow multiple data organizations to coexist. 
For example, let us define four different ways to organize data: 1) by maturity, e.g., raw data \textit{vs.} enriched data; 2) by provenance, e.g., Artefacts and Bibracte; 3) by  confidentiality level, e.g., strictly confidential, restricted or public; and 4) by year of creation.

\begin{figure*}[hbt]
\centering
\includegraphics[width=14.5cm]{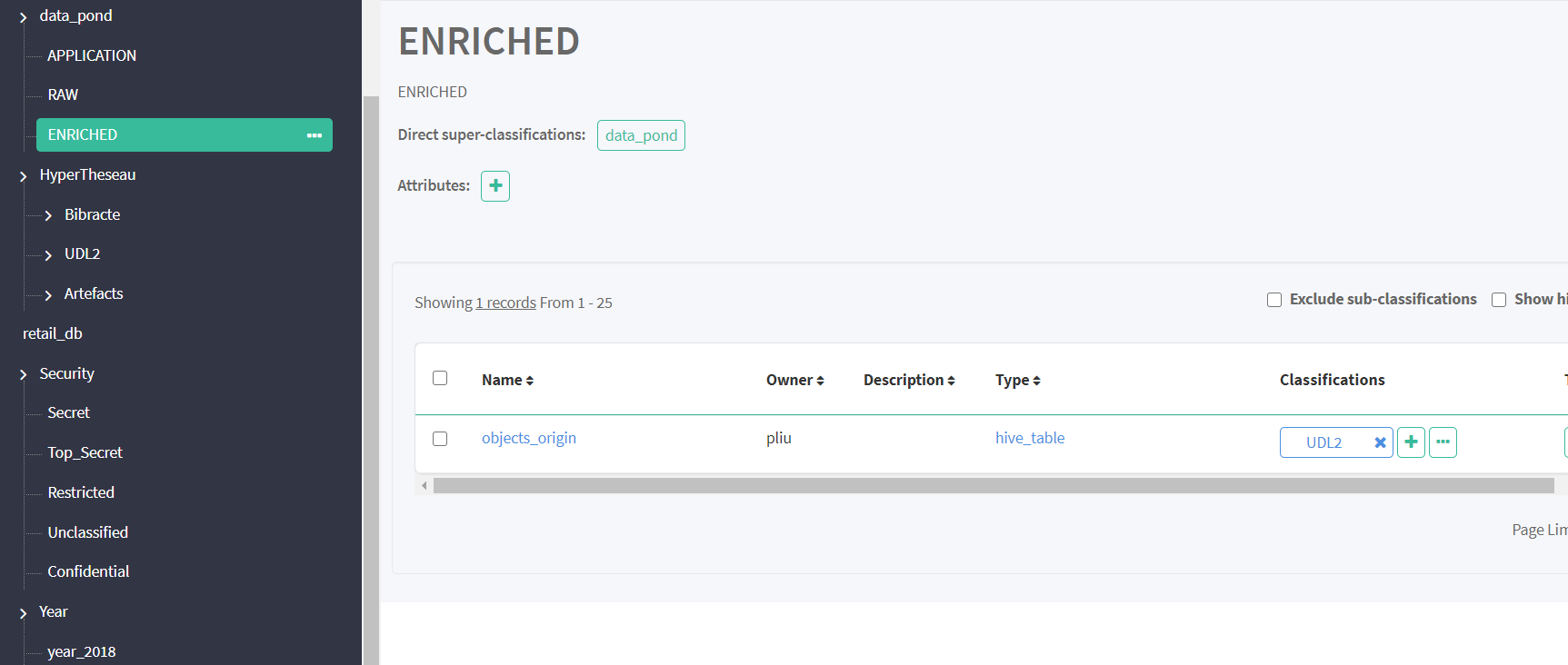}
\caption{Sample Atlas classification}
\label{fig:artefact_data_org}
\end{figure*}

This is achieved through Atlas' classifications, which can group data of the same nature. Moreover, data can be associated with multiple classifications, i.e., be in different data groups at the same time. Figure~\ref{fig:artefact_data_org} shows the implementation of the above data organization in Atlas. We associate table \textit{object\_origin} with four classifications (i.e. enriched, Artefacts, confidential, 2020).  With table \textit{object\_origin} belonging to four classifications, if we want to filter data by maturity, we click on the \textit{enriched} classification to find the table. In sum, classifications allow users to organize data easily and flexibly.

\subsection{Data Indexing and Search through thesauri}

\begin{figure*}[hbt]
\centering
\includegraphics[width=14.5cm]{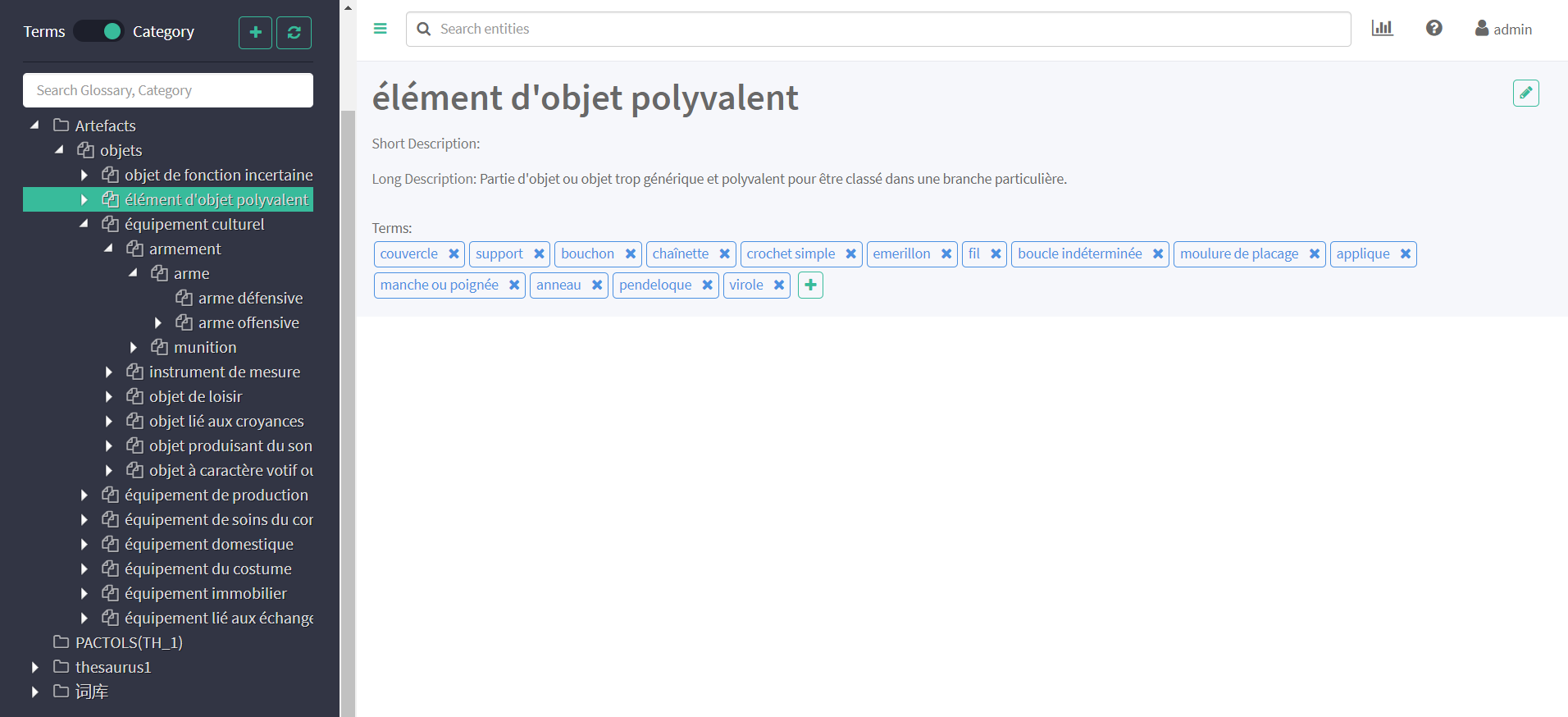}
\caption{Artefacts' thesaurus in Atlas}
\label{fig:artefact_thesaurus}
\end{figure*}

As mentioned in Section~\ref{sec:thesaurus}, thesauri are important metadata for project HyperThesau. As an example, we import a thesaurus provided by the archaeological research facility called \textit{Artefacts} into ArchaeoDAL. This thesaurus is mainly used to index the inventoried archaeological objects of Artefacts. Its basic building blocks are \textit{terms} that can be grouped by \textit{categories} (Figure~\ref{fig:artefact_thesaurus}). 

\begin{figure*}[hbt]
\centering
\includegraphics[width=14.5cm]{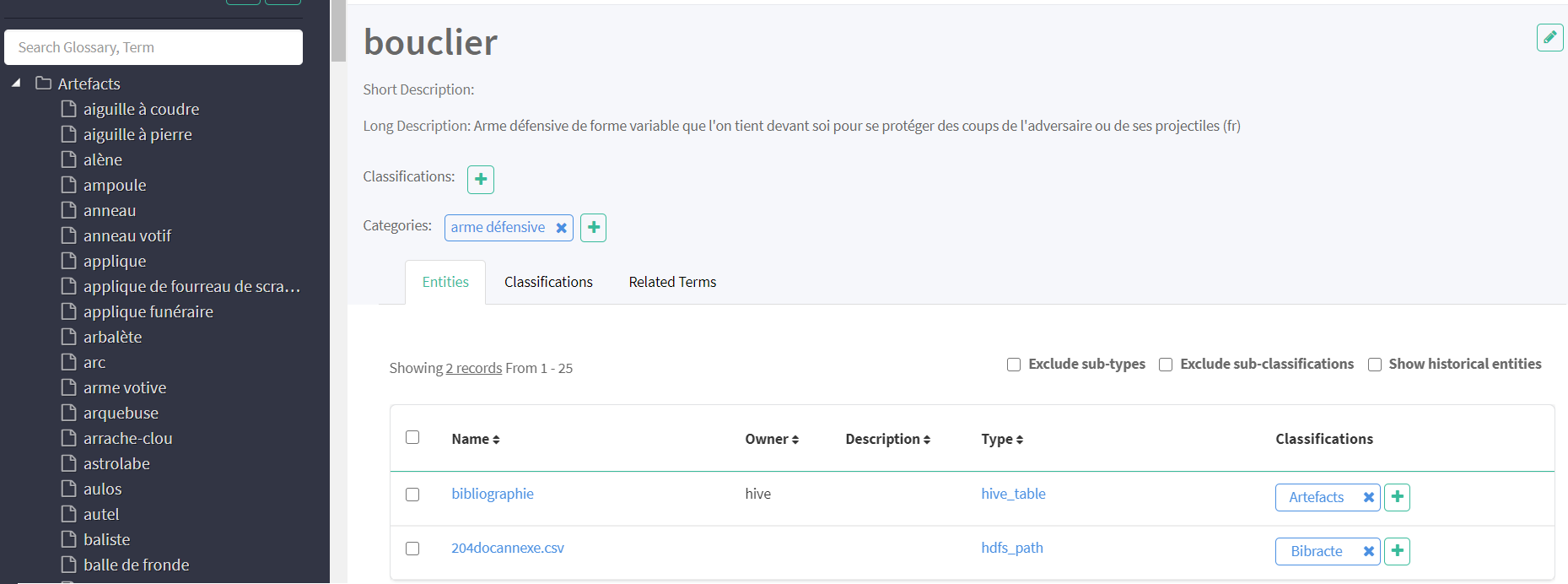}
\caption{Sample data search through Artefacts' thesaurus}
\label{fig:artefact_thesaurus_search}
\end{figure*}

We can associate any data entity with any term. A data entity can be associated with multiple terms. After we index a data entity with a term, we can search data by using the terms of a thesaurus. 
For example, we have a database table called \textit{bibliographie} and a file called \textit{204docannexe.csv} that contains information about shields. Suppose we need to associate these two data entities with the term \textit{bouclier} (shield in French).
After indexing, we can click on the term \textit{bouclier} to find all data associated with this term (Figure~\ref{fig:artefact_thesaurus_search}).

\subsubsection{Data Indexing with Multiple thesauri}
\begin{figure*}[hbt]
\centering
\includegraphics[width=14.5cm]{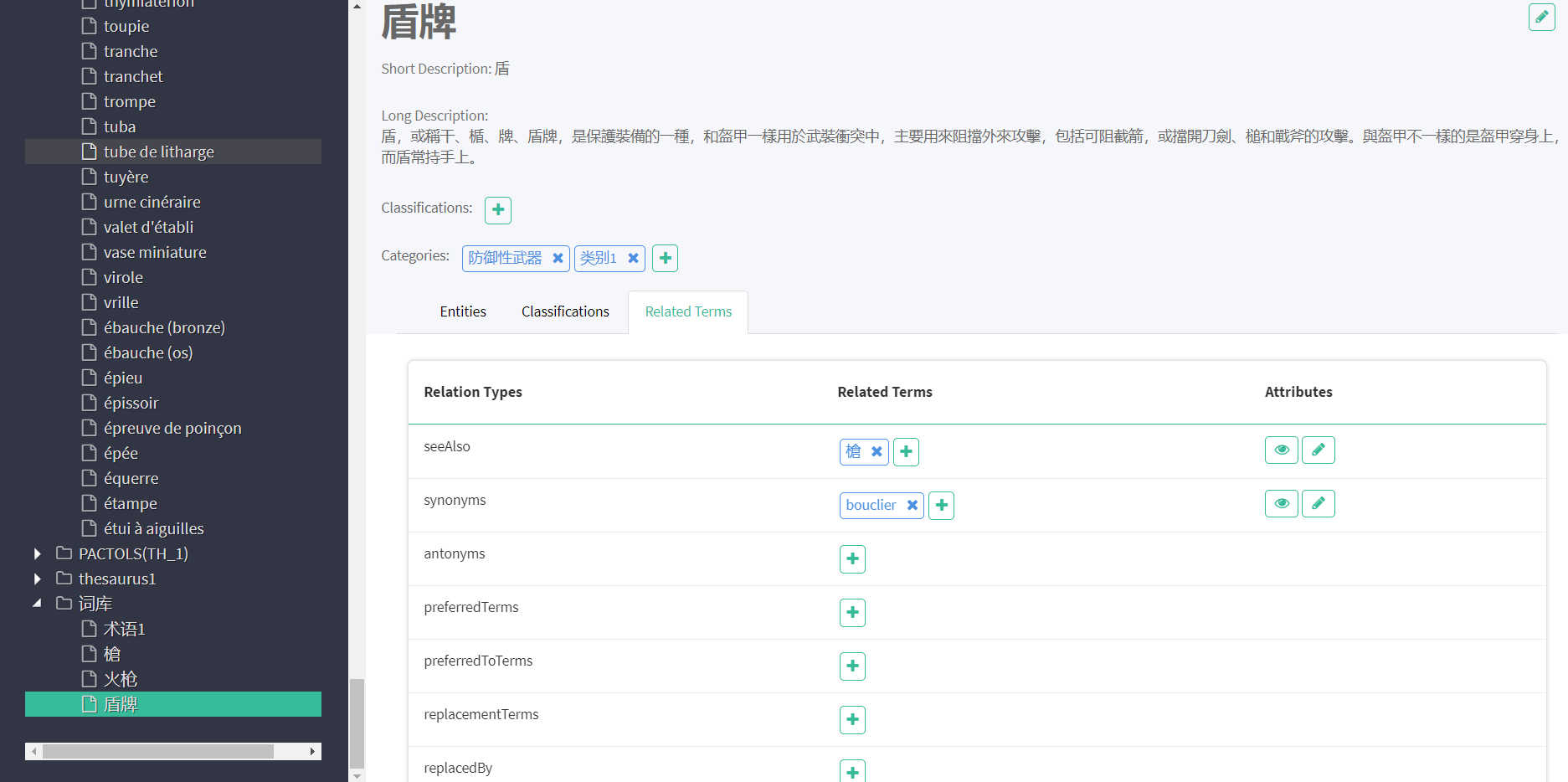}
\caption{Sample term relations in  Artefacts' thesaurus}
\label{fig:artefact_term_relation}
\end{figure*}

One of the biggest challenges of project HyperThesau is that each research facility uses its own thesaurus. Moreover, there is no standard thesaurus. Thus, if we index data with one given thesaurus, archaeologists using another one cannot use ArchaeoDAL.
To overcome this challenge, ArchaeoDAL supports multiple thesauri. Moreover, we can define relations, e.g., synonyms, antonyms, or related terms, between terms of different thesauri. For example, Figure~\ref{fig:artefact_term_relation} shows a term from a Chinese thesaurus that we set as the synonym of the term \textit{bouclier}. As a result, even though the  Chinese term does not relate to any data directly, by using the relations, we can find terms that are linked to actual data. A full video demo of this example can be found online\footnote{\url{https://youtu.be/OmxsLhk24Xo}}.

\section{Conclusion and Future Works}
\label{sec:Conclusion}

In this article, we first introduce the need of archaeologists for software platforms that can host, process and share new, voluminous and heterogeneous archaeological data. Data lakes looking like a viable solution, we examine different existing data lake solutions and conclude that they are not generic, flexible nor complete enough to fulfill project HyperThesau’s requirements. 

As a result, we propose a generic, flexible data lake architecture that covers the full data lifecycle. ArchaeoDAL's architecture is generic because it does not depend on any specific technology. For example, in our current implementation, we use HDFS as the storage layer. Yet, one of our collaborators could easily replace HDFS by Amazon S3. In Section~\ref{sec:flexible-data-org}, we demonstrate how to organize data flexibly, which  many existing solutions~\cite{Ian2019,Inmon2016,Gorelik2019,Alrehamy2015} do not allow. In Section~\ref{sec:flexible-data-org}, we also demonstrate that ArchaeoDAL can gather metadata automatically during the full data lifecycle. Eventually, many features of ArchaeoDAL are very hard to demonstrate in a paper. Thus, we recorded demo videos that are available online\footnote{\url{https://youtube.com/playlist?list=PLrj4IMV47FypKK5WyEd4Oj3-JnfSuU_H1}}.

Archaeologists encounter two major problems while working with ArchaeoDAL. First, to associate data and terms in a thesaurus, domain experts are needed. Moreover, this data-terms matching is a very expensive and time-consuming operation. Thus, we plan to use natural language processing techniques to associate data with a thesaurus automatically, calling domain experts only for \textit{a posteriori} verification.

Second, we handle a lot of images, e.g., aerial photographs and satellite images. It is also very time consuming to detect useful objects in such images. Although some machine learning tasks can already be performed from ArchaeoDAL via Spark-ML, we would like to use deep learning techniques to assist archaeologists in processing images more efficiently.

\bibliographystyle{ACM-Reference-Format}
\bibliography{references.bib}


\begin{thebibliography}{15}


\ifx \showCODEN    \undefined \def \showCODEN     #1{\unskip}     \fi
\ifx \showDOI      \undefined \def \showDOI       #1{#1}\fi
\ifx \showISBNx    \undefined \def \showISBNx     #1{\unskip}     \fi
\ifx \showISBNxiii \undefined \def \showISBNxiii  #1{\unskip}     \fi
\ifx \showISSN     \undefined \def \showISSN      #1{\unskip}     \fi
\ifx \showLCCN     \undefined \def \showLCCN      #1{\unskip}     \fi
\ifx \shownote     \undefined \def \shownote      #1{#1}          \fi
\ifx \showarticletitle \undefined \def \showarticletitle #1{#1}   \fi
\ifx \showURL      \undefined \def \showURL       {\relax}        \fi
\providecommand\bibfield[2]{#2}
\providecommand\bibinfo[2]{#2}
\providecommand\natexlab[1]{#1}
\providecommand\showeprint[2][]{arXiv:#2}

\bibitem[\protect\citeauthoryear{Bird, Campana, Girone, Espinal, McCance, and
  Schovancová}{Bird et~al\mbox{.}}{2019}]%
        {Ian2019}
\bibfield{author}{\bibinfo{person}{Ian Bird}, \bibinfo{person}{Simone Campana},
  \bibinfo{person}{Maria Girone}, \bibinfo{person}{Xavier Espinal},
  \bibinfo{person}{Gavin McCance}, {and} \bibinfo{person}{Jaroslava
  Schovancová}.} \bibinfo{year}{2019}\natexlab{}.
\newblock \showarticletitle{{Architecture and prototype of a WLCG data lake for
  HL-LHC}}.
\newblock \bibinfo{journal}{\emph{EPJ Web of Conferences}}
  \bibinfo{volume}{214} (\bibinfo{year}{2019}), \bibinfo{pages}{04024}.
\newblock
\urldef\tempurl%
\url{https://doi.org/10.1051/epjconf/201921404024}
\showDOI{\tempurl}


\bibitem[\protect\citeauthoryear{Dixon}{Dixon}{2010}]%
        {Dixon2010}
\bibfield{author}{\bibinfo{person}{James Dixon}.}
  \bibinfo{year}{2010}\natexlab{}.
\newblock \bibinfo{title}{{Pentaho, Hadoop, and Data Lakes}}.
\newblock
  \bibinfo{howpublished}{\url{https://jamesdixon.wordpress.com/2010/10/14/pentaho-hadoop-and-data-lakes}}.
\newblock


\bibitem[\protect\citeauthoryear{Fang}{Fang}{2015}]%
        {Fang2015}
\bibfield{author}{\bibinfo{person}{Huang Fang}.}
  \bibinfo{year}{2015}\natexlab{}.
\newblock \showarticletitle{Managing data lakes in big data era: What's a data
  lake and why has it became popular in data management ecosystem}. In
  \bibinfo{booktitle}{\emph{CYBER}}. \bibinfo{publisher}{IEEE},
  \bibinfo{address}{Shenyang, China}, \bibinfo{pages}{820--824}.
\newblock
\urldef\tempurl%
\url{https://doi.org/10.1109/CYBER.2015.7288049}
\showDOI{\tempurl}


\bibitem[\protect\citeauthoryear{Gattiglia}{Gattiglia}{2015}]%
        {Gattiglia2015}
\bibfield{author}{\bibinfo{person}{Gabriele Gattiglia}.}
  \bibinfo{year}{2015}\natexlab{}.
\newblock \showarticletitle{{Think big about data: Archaeology and the Big Data
  challenge}}.
\newblock \bibinfo{journal}{\emph{Archäologische Informationen}}
  \bibinfo{volume}{38} (\bibinfo{year}{2015}).
\newblock
\urldef\tempurl%
\url{https://doi.org/10.11588/ai.2015.1.26155}
\showDOI{\tempurl}


\bibitem[\protect\citeauthoryear{Gorelik}{Gorelik}{2019}]%
        {Gorelik2019}
\bibfield{author}{\bibinfo{person}{Alex Gorelik}.}
  \bibinfo{year}{2019}\natexlab{}.
\newblock \bibinfo{booktitle}{\emph{{Architecting the Data Lake}}}.
\newblock \bibinfo{publisher}{O'Reilly Media}, \bibinfo{address}{North
  Sebastopol, CA, USA}, Chapter~7, \bibinfo{pages}{133--139}.
\newblock
\showISBNx{1491931558}


\bibitem[\protect\citeauthoryear{Inmon}{Inmon}{2016}]%
        {Inmon2016}
\bibfield{author}{\bibinfo{person}{Bill Inmon}.}
  \bibinfo{year}{2016}\natexlab{}.
\newblock \bibinfo{booktitle}{\emph{Data Lake Architecture: Designing the Data
  Lake and Avoiding the Garbage Dump}}.
\newblock \bibinfo{publisher}{Technics Publications},
  \bibinfo{address}{NJ,USA}.
\newblock
\showISBNx{1634621174}


\bibitem[\protect\citeauthoryear{Khatri and Brown}{Khatri and Brown}{2010}]%
        {Vijay2010}
\bibfield{author}{\bibinfo{person}{Vijay Khatri} {and} \bibinfo{person}{Carol
  Brown}.} \bibinfo{year}{2010}\natexlab{}.
\newblock \showarticletitle{Designing data governance}.
\newblock \bibinfo{journal}{\emph{Commun. ACM}}  \bibinfo{volume}{53}
  (\bibinfo{year}{2010}), \bibinfo{pages}{148--152}.
\newblock
\urldef\tempurl%
\url{https://doi.org/10.1145/1629175.1629210}
\showDOI{\tempurl}


\bibitem[\protect\citeauthoryear{McCoy}{McCoy}{2017}]%
        {McCoy2017}
\bibfield{author}{\bibinfo{person}{Mark McCoy}.}
  \bibinfo{year}{2017}\natexlab{}.
\newblock \showarticletitle{{Geospatial Big Data and archaeology: Prospects and
  problems too great to ignore}}.
\newblock \bibinfo{journal}{\emph{Journal of Archaeological Science}}
  \bibinfo{volume}{84} (\bibinfo{year}{2017}), \bibinfo{pages}{74--94}.
\newblock
\urldef\tempurl%
\url{https://doi.org/10.1016/j.jas.2017.06.003}
\showDOI{\tempurl}


\bibitem[\protect\citeauthoryear{Mehmood, Gilman, Cortes, Kostakos, Byrne,
  Valta, Tekes, and Riekki}{Mehmood et~al\mbox{.}}{2019}]%
        {Mehmood2019}
\bibfield{author}{\bibinfo{person}{Hassan Mehmood}, \bibinfo{person}{Ekaterina
  Gilman}, \bibinfo{person}{Marta Cortes}, \bibinfo{person}{Panos Kostakos},
  \bibinfo{person}{Andrew Byrne}, \bibinfo{person}{Katerina Valta},
  \bibinfo{person}{Stavros Tekes}, {and} \bibinfo{person}{Jukka Riekki}.}
  \bibinfo{year}{2019}\natexlab{}.
\newblock \showarticletitle{{Implementing Big Data Lake for Heterogeneous Data
  Sources}}.
\newblock \bibinfo{journal}{\emph{35th IEEE International Conference on Data
  Engineering Workshops (ICDEW)}}  \bibinfo{volume}{2} (\bibinfo{year}{2019}),
  \bibinfo{pages}{37--44}.
\newblock
\urldef\tempurl%
\url{https://doi.org/10.1051/epjconf/201921404024}
\showDOI{\tempurl}


\bibitem[\protect\citeauthoryear{O'Leary}{O'Leary}{2014}]%
        {Oleary2014}
\bibfield{author}{\bibinfo{person}{Daniel O'Leary}.}
  \bibinfo{year}{2014}\natexlab{}.
\newblock \showarticletitle{Embedding AI and Crowdsourcing in the Big Data
  Lake}.
\newblock \bibinfo{journal}{\emph{Intelligent Systems, IEEE}}
  \bibinfo{volume}{29} (\bibinfo{date}{09} \bibinfo{year}{2014}),
  \bibinfo{pages}{70--73}.
\newblock
\urldef\tempurl%
\url{https://doi.org/10.1109/MIS.2014.82}
\showDOI{\tempurl}


\bibitem[\protect\citeauthoryear{Raju, Mital, and Finkelsztein}{Raju
  et~al\mbox{.}}{2018}]%
        {Raju2018}
\bibfield{author}{\bibinfo{person}{Ramakrishna Raju}, \bibinfo{person}{Rohit
  Mital}, {and} \bibinfo{person}{Daniel Finkelsztein}.}
  \bibinfo{year}{2018}\natexlab{}.
\newblock \showarticletitle{Data Lake Architecture for Air Traffic Management}.
  In \bibinfo{booktitle}{\emph{2018 IEEE/AIAA 37th Digital Avionics Systems
  Conference (DASC)}}. \bibinfo{publisher}{IEEE}, \bibinfo{address}{London,
  UK}, \bibinfo{pages}{1--6}.
\newblock
\urldef\tempurl%
\url{https://doi.org/10.1109/DASC.2018.8569361}
\showDOI{\tempurl}


\bibitem[\protect\citeauthoryear{Sawadogo, Scholly, Favre, Ferey, Loudcher, and
  Darmont}{Sawadogo et~al\mbox{.}}{2019}]%
        {Sawadogo2019}
\bibfield{author}{\bibinfo{person}{Pegdwend{\'e}~Nicolas Sawadogo},
  \bibinfo{person}{Etienne Scholly}, \bibinfo{person}{C{\'e}cile Favre},
  \bibinfo{person}{Eric Ferey}, \bibinfo{person}{Sabine Loudcher}, {and}
  \bibinfo{person}{J{\'e}r{\^o}me Darmont}.} \bibinfo{year}{2019}\natexlab{}.
\newblock \showarticletitle{{Metadata Systems for Data Lakes: Models and
  Features}}. In \bibinfo{booktitle}{\emph{{1st International Workshop on BI
  and Big Data Applications (BBIGAP@ADBIS 2019)}}}
  \emph{(\bibinfo{series}{Communications in Computer and Information Science},
  Vol.~\bibinfo{volume}{1064})}. \bibinfo{publisher}{{Springer}},
  \bibinfo{address}{Bled, Slovenia}, \bibinfo{pages}{440--451}.
\newblock
\urldef\tempurl%
\url{https://doi.org/10.1007/978-3-030-30278-8}
\showDOI{\tempurl}


\bibitem[\protect\citeauthoryear{Scholly, Sawadogo, Liu, Alfonso
  Espinosa-Oviedo, Favre, Loudcher, Darmont, and No{\^u}s}{Scholly
  et~al\mbox{.}}{2021}]%
        {dolap2021}
\bibfield{author}{\bibinfo{person}{Etienne Scholly},
  \bibinfo{person}{Pegdwend{\'e}~Nicolas Sawadogo}, \bibinfo{person}{Pengfei
  Liu}, \bibinfo{person}{Javier Alfonso Espinosa-Oviedo},
  \bibinfo{person}{C{\'e}cile Favre}, \bibinfo{person}{Sabine Loudcher},
  \bibinfo{person}{J{\'e}r{\^o}me Darmont}, {and} \bibinfo{person}{Camille
  No{\^u}s}.} \bibinfo{year}{2021}\natexlab{}.
\newblock \showarticletitle{{Coining goldMEDAL: A New Contribution to Data Lake
  Generic Metadata Modeling}}. In \bibinfo{booktitle}{\emph{{23rd International
  Workshop on Design, Optimization, Languages and Analytical Processing of Big
  Data (DOLAP@EDBT/ICDT 2021)}}} \emph{(\bibinfo{series}{CEUR},
  Vol.~\bibinfo{volume}{2840})}. \bibinfo{address}{Nicosia, Cyprus},
  \bibinfo{pages}{31--40}.
\newblock
\urldef\tempurl%
\url{https://hal.archives-ouvertes.fr/hal-03112542}
\showURL{%
\tempurl}


\bibitem[\protect\citeauthoryear{Tomcy and Pankaj}{Tomcy and Pankaj}{2017}]%
        {John2017}
\bibfield{author}{\bibinfo{person}{John Tomcy} {and} \bibinfo{person}{Misra
  Pankaj}.} \bibinfo{year}{2017}\natexlab{}.
\newblock \showarticletitle{Lambda Architecture-driven Data Lake}.
\newblock In \bibinfo{booktitle}{\emph{{Data Lake for Enterprises}}}.
  \bibinfo{publisher}{Packt}, Chapter~2, \bibinfo{pages}{58--77}.
\newblock
\showISBNx{9781787281349}


\bibitem[\protect\citeauthoryear{Walker and Alrehamy}{Walker and
  Alrehamy}{2015}]%
        {Alrehamy2015}
\bibfield{author}{\bibinfo{person}{Coral Walker} {and}
  \bibinfo{person}{Hassan~H. Alrehamy}.} \bibinfo{year}{2015}\natexlab{}.
\newblock \showarticletitle{{Personal Data Lake with Data Gravity Pull}}. In
  \bibinfo{booktitle}{\emph{5th {IEEE} International Conference on Big Data and
  Cloud Computing (BDCloud)}}. \bibinfo{publisher}{IEEE},
  \bibinfo{address}{Dalian, China}, \bibinfo{pages}{160--167}.
\newblock
\urldef\tempurl%
\url{https://doi.org/10.1109/BDCloud.2015.62}
\showDOI{\tempurl}


\end{thebibliography}

\end{document}